\documentclass[twocolumn,superscriptaddress,prl]{revtex4-1}
\usepackage{epsfig}
\usepackage{bm}
\usepackage{fancybox} 
\renewcommand{\thefootnote}{\fnsymbol{footnote}}

\begin{document}

\title{Microscope for X-ray orbital angular momentum imaging} 

\author{Yoshiki Kohmura}
\affiliation{RIKEN, SPring-8 Center, 1-1-1, Kouto, Sayo-gun, Sayo-cho, Hyogo 679-5148, Japan}

\author{Kei Sawada}
\affiliation{RIKEN, SPring-8 Center, 1-1-1, Kouto, Sayo-gun, Sayo-cho, Hyogo 679-5148, Japan}

\author{Masaichiro Mizumaki}
\affiliation{Japan Synchrotron Radiation Research Institute, 1-1-1 Kouto, Sayo, Hyogo, 679-5198, Japan}

\author{Kenji Ohwada}
\affiliation{National Institutes for Quantum and Radiological Science and Technology (QST),  1-1-1, Kouto, Sayo-gun, Sayo-cho, Hyogo 679-5148, Japan}

\author{Tetsuya Ishikawa}
\affiliation{RIKEN, SPring-8 Center, 1-1-1, Kouto, Sayo-gun, Sayo-cho, Hyogo 679-5148, Japan}

\begin{abstract}
Orbital angular momentum (OAM) of photons is carried upon the wave front of
an optical vortex and is important in physics research 
due to its fundamental degree of freedom. 
As for the interaction with materials, the optical OAM was shown to be
 transferred to the valence electron based on modified selection rules
where a single ion is carefully aligned to the center of a vortex\cite{Schmiegelow:2016}. 
We here demonstrate an elaborate way of extracting the distributed OAMs 
in the two-dimensional x-ray wave field at the exit-face of the specimen 
by a vorticity-sensitive microscopy, which detects vorticity 
in the reciprocal-space wave field downstream of the optical vortices.
This method enables us to measure the distributed topological charges 
and OAMs by a single image with the wide field of view over 300 $\mu$m.
Among various wavelength, the research of x-ray OAM is especially interesting 
since it could be closely related to the predicted OAM-induced x-ray dichroism\cite{Veenendaal:2007}. 

\end{abstract}

\maketitle
\medskip


The intensity profile with a central zero, around an optical vortex, is elaborately utilized in
a modern super-resolution microscope\cite{Hell:1994,Hell:2003}
and in nano-fabrication well under the scale of the diffraction limit\cite{Kohmura:2018, Toyoda:2012, Ambrosio:2012,Toyoda:2013, Takahashi:2016}. 
On the other hand, Orbital Angular Momentum (OAM), carried by an optical vortex\cite{Padgett:2004,Zurch:2012, Peele:2002, Taira:2017},
is accelerating its significance recently.
Since quantized OAM number ranges between minus and plus infinity, 
OAM would be useful for enlarging the information channels\cite{Zhou:2016} 
 in a remarked contrast with the spin angular momentum number of $\pm 1$.

To detect OAM of photons, a vorticity-sensitive measurement of the wave field, e.g. downstream of 
the specimens, is highly required. 
Imaging microscopy, making use of the radial Hilbert transform (RHT),
is best suited for this purpose due to its sensitivity to the gradient of complex amplitude,
phase and of amplitude, at the exit-face of specimens\cite{Juchtmans:2016}.  
We hereafter show our experiment using the RHT which enabled us for the first time
to extract the distributed OAMs and to discriminate their topological charge 
at the exit-face of specimens.

The RHT was invented to operate on the input complex amplitudes of 
two dimensional complex data in a similar manner as the Hilbert transform (HT) for one dimensional data. Three operations on the input complex amplitudes, Fourier transform, 
multiplication of a phase filter in the reciprocal space and inverse Fourier transform, results in complex amplitudes     
that equal to the convolution of the input complex amplitudes and the inverse Fourier transform of the phase filter.
\begin{figure}[ht]
\centering\includegraphics[width=0.80\columnwidth]{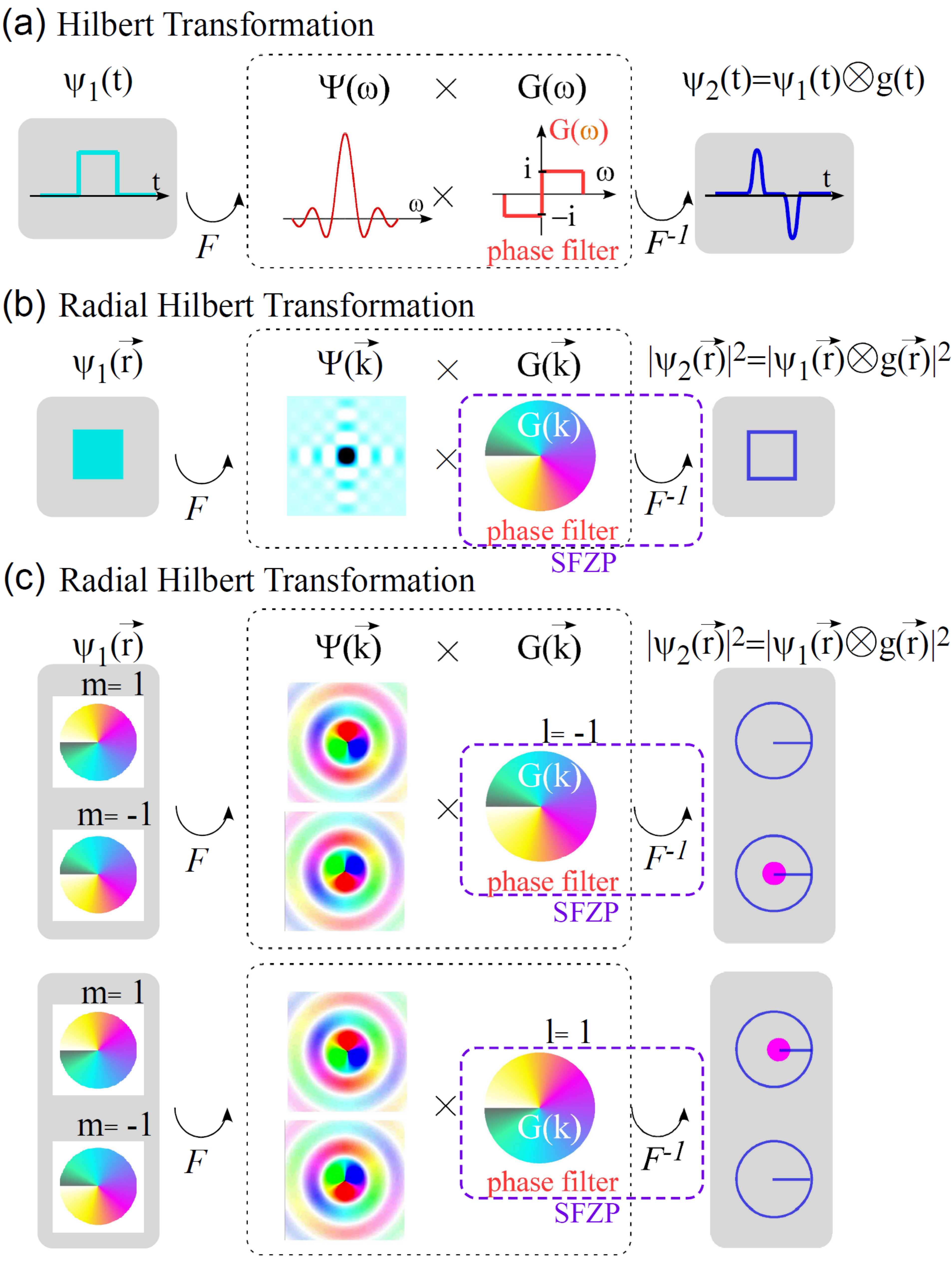} 
\protect\caption{\label{fig:1}
(a) The filter $G(\omega)$ in HT is a step function of $\pi$ in angular frequency space. $g(t)$ is inverse Fourier transform of $G(\omega)$.
(b) The filter in RHT is a spiral phase filter with jump of 2$l\pi$ 
in wave-number space. $g(\vec{r})$ is inverse Fourier transform of $G(\vec{k})$.
Edge enhancement is observed.  
The function of Spiral Fresnel zone plate (SFZP) is indicated by the dotted box in purple. 
(c) The spiral phase anomalies with jump of 2$m\pi$ is discriminated
in RHT when $m$=$l$ is realized. 
Amplitude \& phase are expressed by brightness \& hue, respectively. 
}
\end{figure}
In the RHT, a spiral phase filter with the jump of $2l \pi$ is used as 
the phase filter in the wave-number space, where $l$ is a quantum number.
This transform produces an edge enhanced image of the complex transmissivity 
of the specimen [Fig.~1(a)]. 
The RHT is sensitive to the large gradient existing in the distribution of complex amplitude.

In the RHT, the phase jump of $2l \pi$ is preserved during 
the inverse Fourier transform on the spiral phase filter as shown in Fig.1~(a, b). 
This results in interesting sorting possibilities of objects having spiral phase anomalies, 
e.g. with the exit-face wave field of $\psi_{1}(\vec{r}) \propto \mbox{e}^{im\phi}$ 
having a phase jump of $2m \pi$ where $m$ is a quantum number. 
The complex wave field at the reciprocal space is formed by the Fourier transform 
of the exit-face wave field [$\Psi(\vec{k})$, middle of Fig.~1(c)] which is
multiplied by passing through the spiral phase filter
with the complex transmissivity of $G(\vec{k}) = \mbox{e}^{il\phi_{k}}$, 
where $\phi_{k}$ is the azimuthal angle in the reciprocal space.
Our new theoretical analysis revealed that 
the output complex wave field, $\Psi_{ml}$, formed by the RHT 
shows the intensity enhancements only when the quantum numbers of $m$ 
and $l$ coincide (expressed by $\delta_{ml}$) using a single image, 
which also has position dependences (expressed by $\Psi_{ml}(\vec{r})$)
in the following equation [see Fig.~1(b)],
\begin{equation}
\Psi_{ml}(\vec{r})=\it{F}^{-1} [\mbox{$\Psi(\vec{k}$}) \mbox{e}^{il\phi_{k}}] = \delta_{ml} \mbox{$\Psi_{ml}$}(\mbox{$\vec{r}$}). 
\end{equation}
Note that these equations are for the pure spiral phase filter and the objects with the pure spiral phase anomaly, 
while the neglected effect due to the distribution of attenuation will be reduced by
using low-Z element for these structures.  
We moved on to show the possibility of determining the quantum number of a specimen 
using a single image while the previous theory suggested the measurability of
the rotation of probability current density of the exit wave front using two images\cite{Juchtmans:2016}.
In the present experiment, a Spiral Fresnel zone plate (SFZP)\cite{Sakdinawat:2007} was utilized
to function both as the lens for performing the inverse Fourier transform
and as a spiral phase filter\cite{principle}. 
In this paper, we verify for the first time that the RHT method 
is powerful to sort and analyze the OAM distribution formed by the specimens.

\begin{figure}[bh]
\centering\includegraphics[width=1.0\columnwidth]{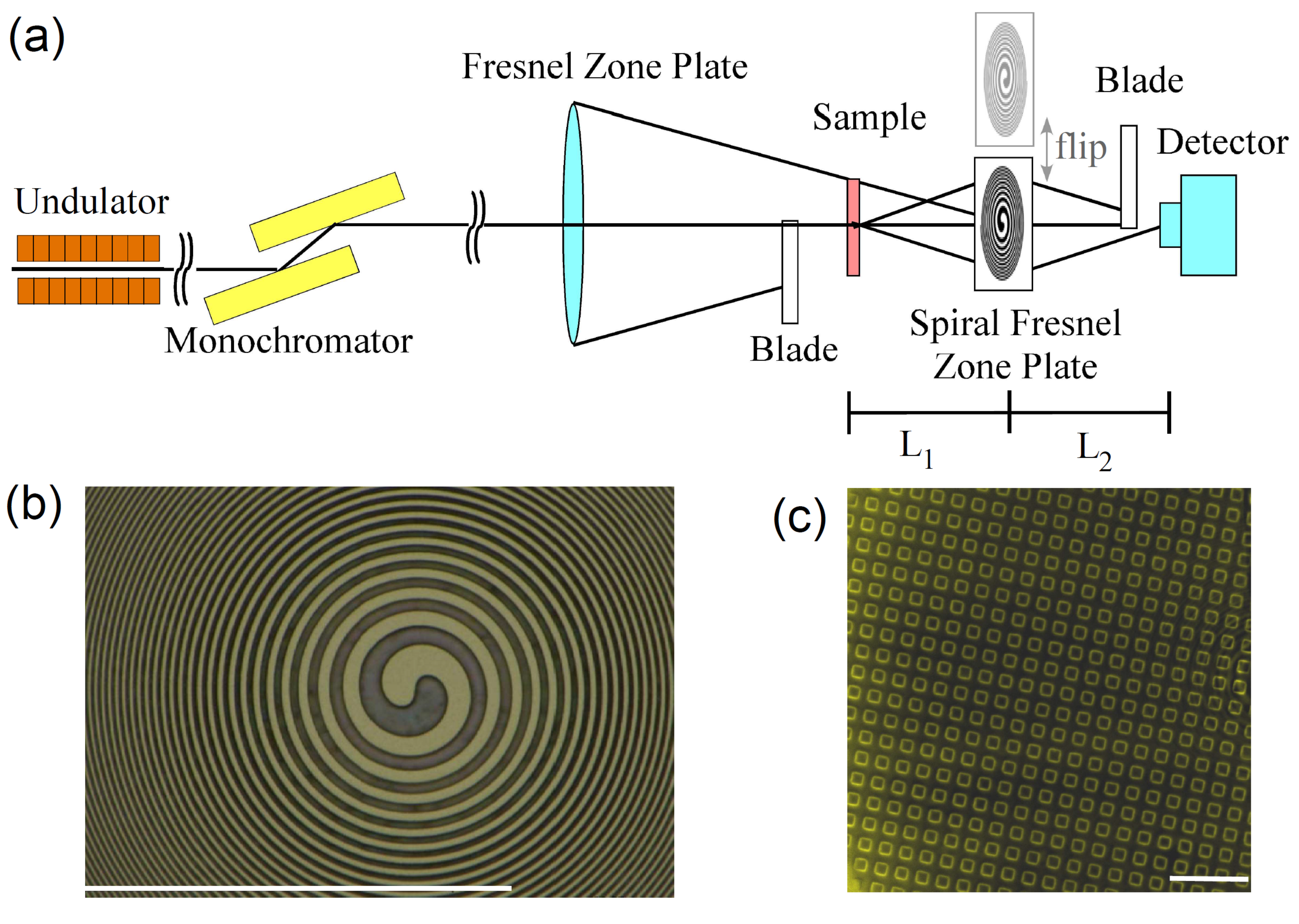} 
\protect\caption{\label{fig:2}
(a) Setup of topological charge microscope with an objective SFZP lens placed at the focal plane of the FZP.
The magnification factor of this microscope is 1.09.
(b) Scanning Electron Microscope image of the SFZP used in the experiment.
The scale bar corresponds to 100 $\mu$m.
(c) Edge enhanced image of copper mesh, without hosting any finite topological charge, 
using the topological charge microscope, with the setup shown in {\bf 2(a)},
exhibiting the rim of the aperture.
The exposure time was 5 seconds.
The scale bar corresponds to 50 $\mu$m.
}
\end{figure} 
The quantum nature of OAM corresponds to (i) a complete circulation 
of the twisting of the vector normal to the wave front that is
Poynting vector, $\vec{S} = \vec{E}^{*}(\vec{r}) \times \vec{H}(\vec{r})+c.c.$
and (ii) the energy current density around an anomaly.
The orientation and the number of this circulation correspond
to the sign and the topological charge of the formed optical vortex.

We construct a RHT microscope using SFZP as the objective lens where
the phase shift aside from the transmissivity of the specimen is observed at the image plane, 
which enables us to discriminate the sign and the topological charge 
of the generated vortices [Fig.~1(c)]. 
We hereafter call this microscope a topological charge microscope.

We can use two topological charge microscope images ($I^{l=1}$ and $I^{l=-1}$) 
obtained with reversed orientations of the SFZP objective lens.
The difference of these two topological charge microscope images exhibits the distribution of
rotation of Poynting vector $\vec{S}$ at the exit-face of 
the specimen\cite{Juchtmans:2016}.
\begin{equation}
I^{l=1}-I^{l=-1} \propto i (\vec{\nabla} \times \vec{S})_{z} \propto i \Bigl[\frac{\partial \psi^{*}}{\partial y} \frac{\partial \psi}{\partial x} - \frac{\partial \psi^{*}}{\partial x} \frac{\partial \psi}{\partial y}\Bigr] = \Omega_{xy},
\end{equation}
where $\Omega_{xy}$ is a Berry curvature term.
Berry curvature, the fictitious magnetic field acting on the wave\cite{Kurosawa:2016}, 
is an important concept applied in various fields of physics such as 
Hall effect, spin Hall effect\cite{Hosten:2008}, optical Hall effect\cite{Onoda:2004}, 
x-ray translation effect inside deformed crystals\cite{Sawada:2006, 
Kohmura:2010, Kohmura:2013, Takei:2016}. 

\begin{figure}[hb]
\centering\includegraphics[width=1.0\columnwidth]{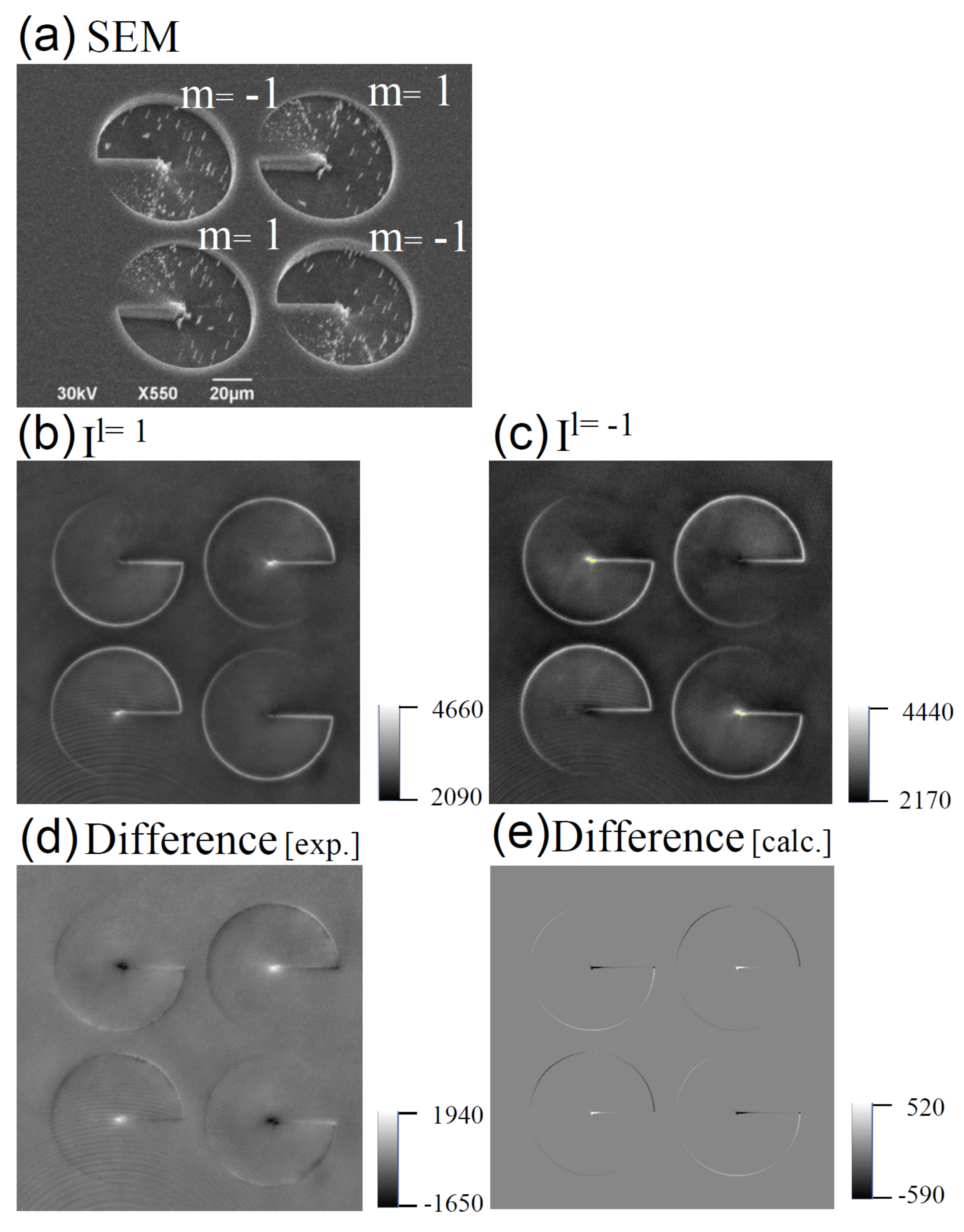} 
\protect\caption{\label{fig:3}
(a) Image of multiple spiral phase filters on the corners of a square
by Scanning Electron Microscope (SEM) with the magnification factor of M=550. 
Topological charge microscope images obtained with the SFZP objective lens 
having the quantum OAM number of (b) $l$=1 (anti-clockwise) and 
(c) $l$=-1 (clockwise from downstream)
and the exposure time of 5 seconds.
(d) Differential microscope image ($I^{l=1}$-$I^{l=-1}$) 
of two images in {\bf 3(b)} and {\bf 3(c)}.  
(e) Calculated differential microscope image using eq. (2). 
}
\end{figure}
X-ray microscopy experiment was performed at Experimental Hutch 2 
along BL29XUL\cite{Tamasaku:2001}, an undulator beamline of SPring-8 facility in Japan,
 using 7.71 keV x-rays from a double crystal monochromator.
The higher order reflection light of monochromator was reduced by a pair of total reflection 
mirrors. The gap of the undulator source was set to 12.207 mm  
to equalize the peak intensity of 1st order harmonics to 7.71 keV. 
The x-ray images have been taken with an indirect x-ray image sensor\cite{Kameshima:2016, detect}. 
An objective SFZP lens was placed at the x-ray focal plane of
the upstream Fresnel Zone Plate (FZP with focal length of 1.72 m) 
while the x-rays transmitted through the specimen 
between FZP and SFZP [Fig.~2(a)]. 
The magnification ratio of the microscope was set to be 1.09
by selecting the distances between sample-SFZP and SFZP-detector
to be 1.40 m and 1.52 m, around twice the focal lengths of SFZP of 0.73 m\cite{spec}.
The zone depth of 1.84 $\mu$m was chosen for 
the tantalum FZP to realize $\pi$-phase shift at 7.71 keV. 
Using this setup, the wave field that corresponds to Fourier transform 
of the specimen's complex transmissivity is irradiated on the SFZP 
plane\cite{Goodman:1996, illuminate}.
This enables us to observe objects with a wide range of spatial length 
scales using a topological charge microscope [Fig.~1(b,c)].
Two blades were placed in front of the specimen and of the imaging detector 
so that the contributions from the other unnecessary orders of light from SFZP, 
aside from 1$^{\mbox{st}}$ order, are excluded in the setup of Fig.~2(a). 
The field of view was reduced to approximately one half by this order selection of the light,
approximately equalizing to the radius of SFZP which is around 324 $\mu$m. 

First, a copper mesh with the pitch of 12.5 $\mu$m, which contains 
no phase anomalies or nor finite topological charge,
was observed using the topological charge microscope. 
As shown in Fig.~1(b), only an edge enhaced image with the bright open squares 
at the rims of the mesh aperture was observed [Fig~2(c)]. 

We then move on to observe objects that yields OAMs or finite topological charges
at the exit-face wave field using the topological charge and its differential microscope. 
For this purpose, we fabricated a specimen having four spiral phase filters 
at the corners of a square on a silicon substrate\cite{specimen} with the maximum depth 
of 19.5 $\mu$m, corresponding to the phase shift of $2 \pi$ for 7.71 keV x-rays.
The radii of spiral phase filters and the distances between the neighbors 
were set to be 34 $\mu$m and 80 $\mu$m, respectively.
Spiral orientation of the phase filters (specimen) was defined according 
to the phase advance in a clockwise ($m$=-1) 
or in an anti-clockwise ($m$=1) orientations from downstream. 
This specimen was placed towards the downstream and 
the x-ray vortices with reversed OAM were realized among the neighbouring 
spiral phase filter [Fig.~3(a)].

The images of this specimen were obtained for 
two setting of the SFZP objective lens [Fig.~2(a)], one in an anti-clockwise ($l=1$) 
and the other in a clockwise ($l=-1$) orientation from downstream,  
where $l$ corresponds to the topological charge of the SFZP [see Figs.~3(b) and ~3(c)]. 
The result clearly indicates that the bright spots are formed at the centers
of the spiral phase filters (specimen) when the quantum numbers of 
the spiral phase filters and SFZP coincides ($m = l$) and are not 
when they are different ($m = -l$). 
The results shown in Figs.~3(b) and ~3(c) agree well with eq. (1). 
The attenuation through these structures was not taken into account in eq.(1), 
but this effect was not significant as shown in these figures. 
The differential image obtained from these two images, which visualizes the distribution of
Berry curvature at the exit-face of the specimen, is shown in Fig.~3(d).
Remarkable structures at the centers and at the rims of 
the spiral phase filters were exhibited which 
were also clearly visible in the calculated differential image shown in Fig.~3(e), derived from eq. (2). 
Note that Fig.~3(e) was calculated by taking account of the attenuation 
through the structures of spiral phase filters in the specimen.

The developed microscope, which visualizes the spatial distributions of OAM from one or two images,
is free from the extremely precise alignment.
The present microscope will open the door of various OAM related researches  
e.g., x-ray OAM-induced dichroism\cite{Veenendaal:2007} or further modification of the selection 
rules that govern the transfer of OAM between valence electrons and photons\cite{Schmiegelow:2016} and so on. 
The chirality research\cite{Brullot:2016} in atomic resolution would be also accelerated.

\subsection{Acknowledgments}
\

Experiments at SPring-8 BL29XU have been performed with the approval of RIKEN 
(under proposal numbers of 20170065). 
Y. K. express a sincere gratitude to Drs. L. Szyrwiel and 
A. Takeuchi for discussions.


\begin{thebibliography}{99}
\bibitem{spec}
The utilized FZP and SFZP were fabricated by the Shinko-Seiki Corp. (Kobe, Japan). 
The specifications of FZP and SFZP were the diameter of 2$r_{N}$= 664 $\mu$m 
\& 648 $\mu$m, the first zone radius of $r_{1}$=16.6 $\mu$m \& 10.8 $\mu$m, 
the outermost zone width of $\Delta r_{N}$= 0.415 $\mu$m \& 0.18 $\mu$m,
number of zones of 400 \& 900, respectively.  
The 7.71keV x-ray focal lengths are 1.72 m and 0.73 m, respectively. 
These ZPs were formed on a membrane on the 500 $\mu$mt silicon substrate, 
back edged to have the thickness of $\simeq 80~\mu$m.
\bibitem{specimen}
The specimen with four spiral phase filters were fabricated 
by the Shinko-Seiki Corp. (Kobe, Japan) on a membrane
formed on a silicon substrate
back edged to have the thickness of $\simeq 80~\mu$m.
\end{thebibliography}
\end{document}